\documentclass[aps,pra,superscriptaddress,nofootinbib,showpacs,doublespacing,preprint]{revtex4-1}

\usepackage{amsfonts}
\usepackage{amsmath}
\usepackage{graphicx}
\usepackage[sort&compress]{natbib}
\usepackage{hyperref}
\usepackage{braket}
\usepackage{overpic}
\usepackage{tikz}
\usetikzlibrary{matrix}
\usepackage{mathrsfs}
\usepackage{amssymb,latexsym}
\usepackage{graphicx}

\begin{document}
\title{Fine structure of the exciton electroabsorption in semiconductor superlattices}
\author{ B.~S.~Monozon}
\affiliation{Physics Department, Marine Technical University, 3 Lotsmanskaya Str.,\\
190008 St.Petersburg, Russia}
\author{P.~Schmelcher}
\affiliation{Zentrum f\"ur Optische Quantentechnologien, Universit\"{a}t Hamburg, \\ Luruper Chaussee 149, 22761 Hamburg, Germany}
\affiliation{The Hamburg Centre for Ultrafast Imaging, Universit\"{a}t Hamburg,~\\ Luruper Chaussee 149, 22761 Hamburg, Germany}
\date{\today}

\begin{abstract}
Wannier-Mott excitons in a semiconductor layered superlattice (SL) of period
much smaller than the 2D exciton Bohr radius in the presence
of a longitudinal external dc electric field directed parallel to the SL axis
are investigated analytically. The exciton states and the
optical absorption coefficient are derived in the tight-binding
and adiabatic approximations. Strong and weak electric fields providing
spatially localized and extended electron and hole states, respectively,
are studied. The dependencies of the exciton states and the exciton absorption
spectrum on the SL parameters and the electric field strength are presented
in an explicit form. We focus on the fine structure of the ground
quasi-2D exciton level formed by the series of closely spaced energy
levels adjacent for higher frequencies. These levels are related to the
adiabatically
slow relative exciton longitudinal motion governed by the potential formed by
the in-plane exciton state. It is shown that the external electric fields compress the
fine structure energy levels, decrease the intensities of the corresponding
optical peaks and increase the exciton binding energy. A possible experimental
study of the fine structure of the exciton electroabsorption is discussed.
\end{abstract}

\maketitle

\section{Introduction}\label{S:intro}

In recent paper of Suris \cite{suris} the
effect of the centre of mass motion
of the Wannier-Mott exciton
on its binding energy in a layered semiconductor
superlattices (SL) has been studied analytically. The
SL period was assumed to be much smaller than the 2D excitonic
Bohr radius. As a result the electron and hole motions
decompose into two parts: the fast longitudinal motion parallel
to the SL axis and the adiabatically slow transverse in-plane
motion governed by the SL potential and the 2D quasi-Coulomb
exciton field, respectively. The fine structure of each 2D
exciton energy level occuring for adjacent but higher energies
was found to occur.
This energy group consists of closely spaced satellite
energy levels which in turn relate to the adiabatically
slow longitudinal relative
exciton motion in the triangular quantum well. This well is
formed by the
quasi-uniform electric field caused by the corresponding
in-plane exciton state. A similar structure was studied
originally in the pioneering work of Kohn and Luttinger
\cite{kohnlutt} when considering the donor states in the
Ge and Si bulk crystals with extremely anisotropic
isoenergetic surfaces. The effect of the longitudinal
uniform external dc electric fields on the exciton states and the exciton
optical absorption spectrum in the semiconductor SL
associated with the transitions to these satellite
states are in this context certainly of relevance.

In the present paper this effect is investigated
for the case that the period of the SL is much smaller than the
2D exciton Bohr radius. In the effective mass approximation
the overlapping exciton wave function is expanded over the
SL Wannier functions that in turn allows us using the
tight-binding and the adiabatic approximations to calculate
analytically the exciton states in the SL subject to the longitudinal
external electric field. The dependencies of the exciton states
and the exciton absorption coefficient on the SL parameters
(period and minibands widths) and on the external
electric field strength
are presented explicitly. The regime of a strong electric field
providing the Wannier-Stark spatial localization
of the carriers and the
more interesting regime of
weak electric field
which destroy the exciton fine structure,
generate the extended longitudinal electron and hole states
and increase the 2D exciton binding energies are considered.
In conclusion we discuss the applicability
conditions of the obtained results and estimate the expected
experimental values.

The paper is organized as follows. In Section 2
in the tight-binding and adiabatic approximations
the equation describing the exciton in the SL in the
presence of the longitudinal external dc electric fields
is derived. The exciton energies and wave functions
are presented in an explicit form in Section 3.
In Section 4 we calculate the exciton absorption
coefficient and trace its dependencies on
the SL parameters and on the external electric field strengths
as well as discuss the applicability of the
obtained results. We also estimate the expected
experimental values. Section 5 contains
the conclusions.

\section{General approach}\label{S:gen}

We consider a Wannier-Mott exciton in a semiconductor SL
subject to an uniform external electric field $\vec{F}$
directed parallel to the SL $z$-axis. The semiconductor
energy bands are taken to be parabolic, nondegenerate,
spherically symmetric and separated by a wide energy gap $E_g$.
In the effective mass approximation the envelope wave
function $\Psi$ of the exciton consisting of the interacting
electron $(e)$ and hole $(h)$ with the effective mass $m_j$,
charges $e_j~(e_e = -e_h = - e) $ and positions
$\vec{r}_j(\vec{\rho}_j ,z_j)~j=e,h$ obeys the equation

\begin{equation}\label{E:basic}
\left\{\sum_{j=e,h}
\left[-\frac{\hbar^2}{2m_j}\vec{\nabla}_j^2 +V_j (z_j) +e_j Fz_j\right]
+ U(\vec{r}_e - \vec{r}_h)\right\}\Psi(\vec{r}_e, \vec{r}_h)
=E\Psi(\vec{r}_e, \vec{r}_h).
\end{equation}
In eq. (\ref{E:basic})

\begin{equation}\label{E:coulomb}
U(\vec{\rho}_e - \vec{\rho}_h, z_e -z_h)=
-\frac{e^2}{4\pi\varepsilon_0\epsilon \sqrt{(\vec{\rho}_e - \vec{\rho}_h )^2 + (z_e -z_h)^2}}
\end{equation}
is the Coulomb potential of the electron-hole attraction in the
semiconductors with the dielectric constant
$\epsilon$, $E$ is the total exciton energy,
$V_j(z_j)= V_j(z_j + nd),~n = 0,1,\ldots N $
are the periodic model potentials of the SL
formed by a large number $N\gg 1$ of quantum wells
of width $d$, separated by $\delta$-function type
potential barriers (see Ref. \cite{zhil92}).
The chosen model correlates well with the nearest neighbor
tight-binding
approximation for the interwell tunneling and enables us
to perform calculations in an explicit form. Since
we follow the original approach to the problem presented
in details in Ref. \cite{suris}, only a brief outline of the
calculations will be provided.

First we assume that the energies of the size-quantization
$b_j$ considerably exceed the exciton Rydberg constant
$Ry$, determined by the exciton Bohr radius $a_0$,
the miniband widths $\Delta_j$ and the distance $eFd$
between the Wannier-Stark (W-S) energy levels, i.e.

\begin{equation}\label{E:adiab}
\Delta_j,\quad Ry,\quad eFd~ \ll ~b_j,
\end{equation}
where

$$
b_j = \frac{\hbar^2 \pi^2}{2m_j d^2};\quad Ry=\frac{\hbar^2}{2\mu a_0^2};\quad
a_0=\frac{4\pi\varepsilon_0\epsilon \hbar^2}{\mu e^2};\quad
\mu^{-1} =m_e^{-1} + m_h^{-1};~j=e,h.
$$

The imposed conditions (\ref{E:adiab}) decompose the particle motion
into two components: the fast longitudinal parallel to the SL $z$-axis
and the slow transverse in-plane $\vec{\rho}$-motions governed,
respectively, by the SL
potentials $V_j (z_j)$ and the electric fields $F$, and the
exciton attraction (\ref{E:coulomb}). This allows us
to employ the approximation of isolated, namely, ground minibands
and expand the exciton function $\Psi$ over the orthonormalized
basis set of the SL Wannier functions $w(z_j - n_j d)$

\begin{equation}\label{E:expan}
\Psi (\vec{r}_e, \vec{r}_h)=\sum_{n_e,n_h}
w(z_e-dn_e)w(z_h-dn_h)\Phi(z_e,\vec{\rho}_e;z_h,\vec{\rho}_h),
\end{equation}
where $z_{e,h}$ are replaced by  
$n_{e,h}=\frac{z_{e,h}}{d}$ in the envelope $\Phi$-function
due to its weak dependence 
on these variables. 
Clearly, the total exciton momentum
with the transverse $\vec{P}$ and longitudinal $Q$ components is
kept constant. The wave function $\Phi$ can be written in the form

\begin{equation}\label{E:Phi}
\Phi(n_e,\vec{\rho}_e;n_h,\vec{\rho}_h)
=\exp\left\{{\rm i}\left[\vec{P}\vec{R}_{\perp} +
\frac{1}{2}Qd(n_e +n_h)-\gamma d n \right]\right\}\chi(n,\vec{\rho}),
\end{equation}
where

\begin{eqnarray}
n &=& n_e-n_h,\quad \gamma(Q)d=
\arctan\left[\frac{\Delta_e -\Delta_h}{\Delta_e + \Delta_h}
\tan \left(\frac{1}{2}Qd\right)\right];
\nonumber\\
\vec{R}_{\perp} &=& \frac{m_e \vec{\rho}_e +m_h \vec{\rho}_h}{M},~
\mbox{and }~\vec{\rho} = \vec{\rho}_e - \vec{\rho}_h,~ M=m_e + m_h,
\nonumber
\end{eqnarray}
are the transverse centre of mass $(\vec{R}_{\perp} )$ and relative
$(\vec{\rho})$ coordinates, respectively.
Substitution of functions (\ref{E:expan}) and (\ref{E:Phi})
into eq. (\ref{E:basic}) and subsequent projection yields
the equation for the function $\chi(n,\vec{\rho})$

\begin{eqnarray}\label{E:relat}
&&-\frac{eFd}{2}\left\{\left(n+\frac{\mathscr{E}}{eFd} -\beta \right)2\chi(n,\vec{\rho})
+ \beta \left[ \chi(n+1,\vec{\rho})+\chi(n-1,\vec{\rho}) \right] \right\}+
\nonumber\\
&&\left[-\frac{\hbar^2}{2\mu}\vec{\nabla}_{\vec{\rho}}^2 + U(n,\vec{\rho})
- W \right]\chi(n,\vec{\rho})=0,
\end{eqnarray}
where

\begin{eqnarray}
&&\beta = \frac{\Delta_{eh}(Q)}{2eFd},~
W = E - \tilde{E}_g -\frac{P^2}{2M} - T(Q)-\mathscr{E},~
\tilde{E}_g = E_g + b_e + b_h,
\nonumber\\
&&U(n,\vec{\rho}) = \big < w(z_e - dn_e)w(z_h - dn_h)\mid U(\vec{\rho}_e - \vec{\rho}_h, z_e -z_h) \mid
w(z_e - dn_e)w(z_h - dn_h)\big>.
\nonumber
\end{eqnarray}
The energy

$$
T(Q)= \frac{1}{2}\left[ \Delta_e + \Delta_h - \Delta_{eh}(Q) \right  ];~
\Delta_{eh}(Q)= \left[(\Delta_e + \Delta_h)^2 -
2\Delta_e \Delta_h(1 - \cos Qd) \right]^{\frac{1}{2}}
$$
is the energy of the centre of mass longitudinal motion.
Here it is reasonable to introduce the parameters

$$
m_{\|}(Q) = \frac{2\hbar^2}{\Delta_{eh}(Q) d^2},~\mbox{and}~
a_F (Q) = \left( \frac{\hbar^2}{2m_{\|}(Q) eF} \right)^{\frac{1}{3}},
$$
which are the reduced longitudinal
effective mass $(m_{\|}(Q))$, determining the relative
$z$-motion of the electron-hole pair and the effective
length $(a_F (Q))$ of the corresponding state.
On solving eq. (\ref{E:relat}) the exciton wave function
$\Psi$ (\ref{E:expan}) and the exciton energy $E$ can be calculated
in principle.

\section{Exciton states}\label{S:States}

\emph{Wannier-Stark (W-S) regime} $a_F (Q) \leq d~ (\beta \leq 1)$
\\
\\
Below we consider sufficiently strong electric fields $F$ providing
$z$-localization of the carriers within several periods and
quantization of its longitudinal states (W-S levels). The
solution to eq. (\ref{E:relat}) can be written in the adiabatic form

\begin{equation}\label{E:rel1}
\chi(n,\vec{\rho}) = R_{p(k)}(n,\vec{\rho})\psi_{\nu} (n),
\end{equation}
where $R_{p(k)}(n,\vec{\rho})$ is the 2D exciton wave function of the
discrete $(p)$ (continuous $(k)$) states \cite{zhil92,monzhil94}
determined by the potential $U(n,\vec{\rho})$ and corresponding to the
energies

\begin{equation}\label{E:coulen}
W_p = - \frac{Ry}{(p+\delta p +\frac{1}{2})^2},~p = 0,1,2,\ldots ,~
\delta p \simeq \frac{2}{3}\frac{d}{a_0} \ll 1,
~\mbox{and}~W_k =\frac{\hbar^2 k^2}{2\mu}.
\end{equation}
Quantum defects $\delta p$ have been calculated earlier in Ref. \cite{monschm05}
in the framework of the chosen model potentials $V_j (z_j)$ \cite{zhil92}.
The solution $\psi_\nu$ in the function $\chi(n,\vec{\rho})$ (\ref{E:rel1})
to the difference equation (\ref{E:relat})
for a vanishing of the term in the curly brackets,
is well known (see for example Ref. \cite{ivch} and references therein)

\begin{equation}\label{E:wsstate}
\psi_{\nu}(n)=J_{-(n+\nu)}(\beta),~
\mathscr{E}_{\nu}=eFd\nu +\frac{1}{2}\Delta_{eh},~\nu = 0,\pm 1,\pm 2,\ldots ,
\end{equation}
where $J_{m}(x)$ and $\mathscr{E}_{\nu}$ are the Bessel functions
and the W-S energy levels, respectively. Eqs. (\ref{E:coulen}), (\ref{E:wsstate}),
(\ref{E:rel1}) and (\ref{E:Phi}) lead to the total exciton energies

\begin{equation}\label{E:exciten}
E_{\nu,p(k)}(\vec{P},Q) =
\tilde{E}_g +\frac{P^2}{2M} + T(Q) +\mathscr{E}_{\nu} + W_{p(k)}
\end{equation}
and the orthonormalized wave functions $\Psi$ (see eq. (\ref{E:expan}))

\begin{equation}\label{E:Psi}
\Psi_{\nu,p(k)}^{(\vec{P},Q)}(z_e,z_h ; \vec{\rho}, \vec{R}_{\perp})=
\frac{{\rm e}^{{\rm i}\vec{P}\vec{R}_{\perp}}}{\sqrt{SN}}
\sum_n{\rm e}^{{\rm i}\left(\frac{1}{2}Q - \gamma \right)dn}
J_{-(n+\nu)}(\beta)R_{p(k)}(n,\vec{\rho})G_n^{(Q)}(z_e,z_h),
\end{equation}
where

$$
G_n^{(Q)}(z_e,z_h)=\sum_{n_{h}} w[z_e - d(n + n_h)  ]
w[z_h - dn_h ]{\rm e}^{{\rm i}Qn_h},
$$
$S$ is the area of the SL layer.
The wave functions $\Psi_{\nu,p(k)}^{(\vec{P},Q)}$ satisfy

$$
\big <\Psi_{\nu,p(k)}^{(\vec{P},Q)}\mid \Psi_{\nu',p'(k')}^{(\vec{P}',Q')}\big>=
\delta_{\vec{P}\vec{P}'}\delta_{QQ'}\delta_{\nu \nu'}\delta_{p(k)p'(k')}.
$$
Eqs. (\ref{E:exciten}) and (\ref{E:Psi}) at $\vec{P}=Q=0$ coincide with those
obtained for the exciton energies and wave functions, respectively, in
Ref. \cite{zhil92}.
\\
\\
\emph{Continuous regime} $a_F (Q) \gg d~ (\beta \gg 1)$
\\
\\
Weak electric fields $F$ generate extended longitudinal
exciton states, for which the continuous limit implying in eq.
(\ref{E:relat})

$$
nd=z,~\chi(n+1, \vec{\rho}) + \chi(n-1, \vec{\rho})-
2\chi(n, \vec{\rho})= d^2\frac{\partial ^2\chi(z, \vec{\rho})}{\partial z^2}
$$
becomes applicable. The function $\chi(z, \vec{\rho})$ obeys the equation

\begin{eqnarray}\label{E:relat1}
&&\left[-\frac{\hbar^2}{2m_{\|}(Q)}\frac{\partial^2}{\partial z^2} -
eFz - \frac{\hbar^2}{2\mu}\vec{\nabla}_{\vec{\rho}}^2 +
U(z,\vec{\rho})-W  \right]\chi(z, \vec{\rho})=0;
\nonumber\\
&&W=E-\tilde{E}_g-\frac{P^2}{2M} - T(Q).
\end{eqnarray}

Since $m_{\|}(Q) \gg \mu$ (see condition (\ref{E:adiab}))
the slow $z$- and fast $\vec{\rho}$-motions are separated adiabatically
which in turn enables us to take
$\chi(z, \vec{\rho})=R_p(\vec{\rho})\psi_p(z),~p=0,1,\ldots $, where the functions
$R_p(\vec{\rho})$ describe the exciton states governed by the 2D
Coulomb potential $U(0,\vec{\rho})\sim \rho^{-1}$ (\ref{E:coulomb})
with the exact Rydberg energies $W_p$ (\ref{E:coulen}) at $\delta p =0$.
Below to be definite we consider the ground 2D state with
$R_0 (\vec{\rho}) = (2\pi)^{-\frac{1}{2}}4a_0^{-1}\exp (-2\rho/a_0)$
and $W_0 = - 4Ry$. For the wave function
$\psi_0 (z)$ we obtain from eq. (\ref{E:relat1})

\begin{equation}\label{E:triangle}
-\frac{\hbar^2}{2m_{\|}(Q)}\psi_0^{''}(z) +
\left[\bar{U}_0(z)- eFz - \mathscr{E}_0 \right]\psi_0 (z)=0;~
\mathscr{E}_0 = W - W_0,
\end{equation}
where

\begin{eqnarray}\label{E:1Dpot}
&&\bar{U}_0(z) =\big<R_0 (\vec{\rho})\mid U(\rho,z) - U(\rho,0)\mid R_0 (\vec{\rho})\big>=
\nonumber\\
&&8Ry\left(|u |+\frac{1}{2}u^2\ln|u|-\frac{1}{3}|u|^3\right);~|u| =\frac{4|z|}{a_0}\ll 1.
\end{eqnarray}

Equation (\ref{E:triangle}) at $F=0$ for the region of small $z$-coordinate
has been considered by Kohn and Luttinger \cite{kohnlutt} and Suris \cite{suris}
for the cases of a bulk crystal and SL, respectively.
The symmetric triangular potential well

$$
\bar{U}_0(z) = eF_0|z|,\quad F_0 =\frac{32Ry}{ea_0}
$$
formed by a quasi-uniform exciton electric field $F_0$
and the potential of the uniform external electric field $F \leq F_0$
form the asymmetric triangular
potential well. This enable us to present the exact solution to eq. (\ref{E:triangle})
in terms of the Airy functions $Ai(x)$ \cite{abram}

\begin{eqnarray}\label{E:psi}
\psi_{0s}(z)=
\left\{
\begin{array}{cl}
Ai\left(\frac{z-z_{+}}{a_{+}}\right);~z\geq 0;\\
Ai\left(\frac{z-z_{-}}{a_{-}}\right);~z\leq 0,
\end{array}
\right.
\end{eqnarray}
where

\begin{eqnarray}
z_{+,-}&=&\frac{\mathscr{E}_{0s}}{e(F_0 \mp F)};~
a_{+,-} = a_{F_0}(Q) \left(1\mp \frac{F}{F_0}\right)^{\frac{1}{3}};~
a_{F_0}(Q) = \left(\frac{\hbar^2}{2m_{\|}(Q)eF_0}\right)^{\frac{1}{3}};~
\nonumber\\
C^2& =& \left(\frac{2\pi^2}{2s+1}\right)^{\frac{1}{3}}\frac{1}{a_{F_0}(Q)}
\left\{\left(\frac{1 + \frac{F}{F_0}}{1 - \frac{F}{F_0}}\right)^{\frac{1}{3}}
 + \left(\frac{1 - \frac{F}{F_0}}{1 + \frac{F}{F_0}}\right)^{\frac{1}{3}} \right\}^{-1}.
\nonumber
\end{eqnarray}

The satellite energies

\begin{equation}\label{E:trien}
\mathscr{E}_{0s}=(12\pi)^{\frac{2}{3}}Ry
 \left( \frac{\mu}{m_{\|}(Q)}\right)^{\frac{1}{3}}
 \left[\left(1- \frac{F^2}{F_0^2}  \right)\left(2s+1  \right)\right]^{\frac{2}{3}};~
s=0,1,2,\ldots
\end{equation}
adjacent to the energy level $W_0$ determine the total exciton energies

\begin{equation}\label{E:exciten1}
E_{0s}(\vec{P},Q) =
\tilde{E}_g +\frac{P^2}{2M} + T(Q) +\mathscr{E}_{0s} + W_0.
\end{equation}
Eqs. (\ref{E:psi}) and (\ref{E:Phi}) lead to the
orthonormalized total exciton wave functions (\ref{E:expan})

\begin{equation}\label{E:Psi1}
\Psi_{0s}^{(\vec{P},Q)}(z_e,z_h ; \vec{\rho}, \vec{R}_{\perp})=
\frac{{\rm e}^{{\rm i}\left(\vec{P}\vec{R}_{\perp} +
QZ -  \frac{1}{2}\frac{\Delta_e -\Delta_h}{\Delta_e +\Delta_h}Qz\right)}}{\sqrt{SNd}}
R_0(\vec{\rho})\psi_{0s}(z),
\end{equation}
with

$$
\big <\Psi_{0s}^{(\vec{P},Q)}\mid \Psi_{0s'}^{(\vec{P}',Q')}\big>=
\delta_{\vec{P}\vec{P}'}\delta_{QQ'}\delta_{ss'}.
$$

\section{Spectrum of the exciton absorption: Results and discussion}\label{S:spectrum}

The relation between the transition rate $\Pi$
and the exciton optical absorption coefficient $\alpha$
is as follows \cite{monzhil94}

\begin{equation}\label{E:wsabs}
\alpha=\frac{n_0 \hbar\omega\Pi}{c\tilde{U}SNd};~
\Pi = \frac{1}{t}\sum_{e,h}
\big|\frac{1}{{\rm i} \hbar}\int_0^t \Psi_0(\vec{r}_e,\vec{r}_h)
\mathscr{P}_{eh}(\tau)\Psi^{*}(\vec{r}_e,\vec{r}_h)
{\rm e}^{{\rm i}\frac{E}{\hbar}}d\vec{r}_e d\vec{r}_h\big|^2,
\end{equation}
where $n_0$ is the refractive index, $c$ is the speed of light,
$\tilde{U}=\varepsilon_0 n_0^2 F_0^2$ is the optical energy
density caused by an oscillating electric field
of frequency $\omega$ and magnitude $F_0$ polarized
parallel to the SL axis. $\sum_{e,h}$ is a sum over all band states
involved in optical transitions and
 $\Psi_0(\vec{r}_e,\vec{r}_h)= \delta (\vec{r}_e - \vec{r}_h)$
 is the wave function of the initial electron-hole state \cite{ell}.
$\Psi(\vec{r}_e,\vec{r}_h)$ and $E$ are the exciton wave
function and energy, respectively,

$$
\mathscr{P}_{eh}(t) = \frac{{\rm i}2\hbar eF_0p_{ehz}}{m_0 E_g}\cos \omega t;~
( E_g \simeq \hbar \omega)
$$
is the operator of allowed electric dipole transitions \cite{weiler},
determined by the free electron mass $m_0$ and momentum $z$-component
matrix element $p_{ehz}$ calculated between the amplitudes
of the Bloch functions of the electron and hole bands.
\\
\\
\emph{Wannier-Stark regime} $a_F (Q) \leq d~ (\beta \leq 1)$
\\
\\
Substituting the exciton wave function $\Psi$ (\ref{E:Psi}) with
$R_{p(k)}(n,\vec{\rho})$ from Ref. \cite{monzhil94} and energy $E$
(\ref{E:exciten}) into eq. (\ref{E:wsabs}) we obtain

\begin{eqnarray}\label{E:spectr1}
\alpha=\sum_{\nu}
\left[\frac{\pi e^2 \hbar}{2\varepsilon_0 n_0 Vm_0c}
\sum_{p=0}^{\infty} f_{\nu p}^{(2\mbox{\tiny D})}
\delta \left(\hbar \omega -\tilde{E}_g - \mathscr{E}_{\nu} +
\frac{Ry}{\left(p+\delta p +\frac{1}{2} \right)^2} \right)\right.
\nonumber \\
+
\left.\alpha^{(0)}J_{-\nu}^2(\beta)\frac{2\mu}{\pi\hbar^2}
\frac{{\rm e}^{\xi_{\nu}(\omega)}}{\cosh \xi_{\nu}(\omega)}
\Theta(\hbar \omega -\tilde{E}_g - \mathscr{E}_{\nu} ) \right],
\end{eqnarray}
where

\begin{eqnarray}\label{E:strength1}
f_{\nu p}^{(2\mbox{\tiny D})} &=& J_{-\nu}^2(\beta)
\frac{4V \hbar\omega |p_{ehz}|^2}
{\pi a_0^2\left(p+\delta p +\frac{1}{2}\right)^3 m_0 d E_g^2};~
\alpha^{(0)} = \frac{2\pi\hbar^2\omega e^2 |p_{ehz}|^2}
{\varepsilon_0 n_0m_0^2 cd E_g^2};~
\nonumber \\
\xi_{\nu}(\omega) &=& \pi \left(\frac{\hbar \omega - \tilde{E}_g - \mathscr{E}_{\nu}}{Ry}   \right)^{-\frac{1}{2}},
\end{eqnarray}
$f_{\nu p}^{(2\mbox{\tiny D})}\, \mbox{is the osillator strength
of the transition to the  }\, \nu p\,~ \mbox{state and}\,
\Theta (x)$ is the Heavyside step function.
\\
\\
\emph{Continuous regime} $a_F (Q) \gg d~ (\beta \gg 1)$
\\
\\
At this stage we take in eq. (\ref{E:wsabs})
eqs. (\ref{E:Psi1}) and (\ref{E:exciten1})
for the wave function $\Psi$ and energy $E$, respectively.
The coefficient of absorption becomes

\begin{equation}\label{E:spectr2}
\alpha = \frac{\pi e^2 \hbar}{2\varepsilon_0 n_0 Vm_0c}
\sum_s f_{0s}\delta
\left(
\hbar \omega - \tilde{E}_g + E_{0s}^{(b)}
\right)
\end{equation}
where

\begin{equation}\label{E:strength2}
f_{0s} = \frac{16V \hbar \omega \mid p_{ehz}  \mid ^2}{\pi m_0 E_g^2 a_0^2 a_{F_0}(0)}
\left(\frac{2}{\pi}\right)^{\frac{2}{3}}\Lambda_s (F)
\end{equation}
with

$$
\Lambda_s (F) = \frac{ \sin^2 \left[\frac{\pi}{3} (2s+1) \left(1+\frac{F}{F_0}\right) + \frac{\pi}{4}\right] }
{ (2s+1)^{\frac{2}{3}} }
\frac{2\left(1-\frac{F}{F_0}   \right)^{\frac{1}{3}} }
{\left(1-\frac{F}{F_0}   \right)^{\frac{2}{3}  }
+\left(1+\frac{F}{F_0}   \right)^{\frac{2}{3}  } }
$$
and

\begin{equation}\label{E:bind}
E_{0s}^{(b)}=4 Ry
\left[
1 - \left(\frac{\mu}{m_{\parallel}(0)}\right)^{\frac{1}{3}}
\left( \frac{3\pi}{2}\right)^{\frac{2}{3}}\Omega_s (F)\right]
\end{equation}
with

$$
\Omega_s (F) = \left[(2s+1) \left(1-\frac{F^2}{F_0^2}\right) \right]^{\frac{2}{3}}
$$
are the oscillator strengths of the transitions to the $0s$ states
$(f_{0s})$ and corresponding binding energies $(E_{0s}^{(b)})$, respectively.
The oscillator strengths $f_{0s}$ scaled with the 2D
oscillator strength $f_{00}^{(2\mbox{\tiny D})}$
calculated from eqs. (\ref{E:strength1}) for
$\beta\ll 1~(J_0 (\beta)=1)$
and the blue shifts $\hbar\omega -(\tilde{E}_g - 4Ry)$
of the ground $s=0$ and excited $s=1,2,3$
satellite exciton peaks scaled with the 2D exciton
binding energy $4Ry$ can be expressed via the functions
$\Lambda_s (F)$ and $\Omega_s (F)$, respectively, by

\begin{equation}\label{E:link}
 \frac{f_{0s}}{f_{00}^{(2\tiny D)}}= \frac{d}{a_0}
 \left(\frac{16 m^{\parallel}(0)}{\pi^2 \mu}\right)^{\frac{1}{3}}
 \Lambda_s (F)
 \end{equation}
and

\begin{equation}\label{E:link1}
 \frac{\hbar\omega- (\tilde{E}_g - 4Ry)}{4Ry}=
 \left(\frac{9\pi^2 \mu}{4 m^{\parallel}(0)}\right)^{\frac{1}{3}}
 \Omega_s (F).
\end{equation}

Since the exciton electroabsorption in
the W-S regime has been discussed in details by
Zhilich \cite{zhil92} we comment only briefly on eq.
(\ref{E:spectr1}) below. The absorption spectrum
consists of a periodic sequence of equidistant blocks
each of them is formed by the Rydberg series of discrete
$p$-peaks and continuous branches $\xi_{\nu}(\omega)$
adjacent to the W-S threshold
$\hbar \omega_{\nu}=\tilde{E}_g + \mathscr{E}_{\nu}~ (T(0)=0)$
from low and high frequencies, respectively.
The distance between the neighboring blocks
$\hbar\Delta \omega = eFd$ considerably exceeds
the energy width $Ry$ of the series. Within the
Rydberg series the ground peak $p=0$ dominates
others, which decrease in intensity as
$\sim (p+\frac{1}{2})^{-3}$ with increasing indices $p$.
In the presence of sufficiently strong electric fields
$(\beta < 1)$ the exciton series corresponding to the threshold
$\hbar \omega_0 = \tilde{E}_g +\frac{1}{2}\Delta_{eh}$
becomes more pronounced, while intensities of others
$\sim \beta ^{2|\nu|}$ decrease with increasing the index of
the subband $\nu$.

For weak electric fields $F$, providing the continuous
regime, the absorption spectrum consists of Rydberg
series formed by the $p$-peaks, each of them has the
fine structure represented by the satellite $s$-lines located
above the energy $\hbar \omega_p =\tilde{E}_g -\frac{4Ry}{(2p+1)^2}$.
In the absence of the electric field $F$
the oscillator strengths of the satellite peaks
$f_{0s}$ related to the ground exciton state $p=0$ are presented
in Fig.1. The intensities of the $0s$ satellite lines
oscillate with increasing the
quantum number $s$. The longitudinal motion with the finite
effective mass $m_{\parallel}(0)$ reduces the binding energy
$E_{0s}^{(b)}$, blue shifts the exciton peaks and causes
the oscillator strengths $f_{0s}$
(\ref{E:strength2}) to be small compared to the corresponding
strictly 2D values $f_{00}^{(2\mbox{\tiny D})}$
(see eq. (\ref{E:final})). For the ground
exciton state $p=s=0$ the correction to the binding energy
$E_{00}^{(b)} - 4Ry$ scaled with the 2D binding energy
$4Ry$ and the ratio of the peak intensities are

\begin{figure}[htbp]
 \begin{center}
\includegraphics[width= .75\linewidth]{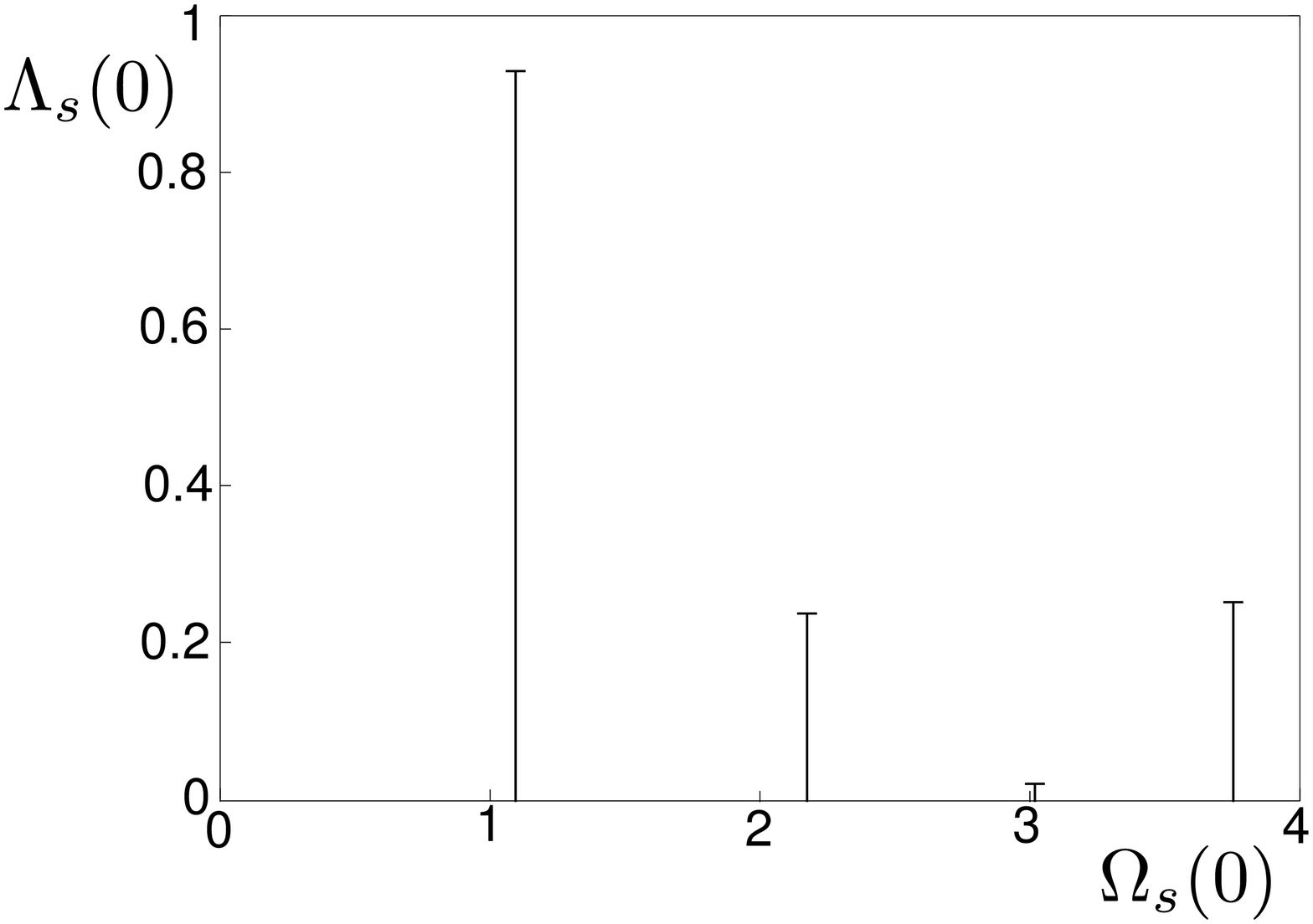}
 \end{center}
 \caption{The oscillator strengths of the satellite exciton
peaks $f_{0s}$ scaled with the 2D
oscillator strength $f_{00}^{(2\mbox{\tiny D})}$
calculated from eqs. (\ref{E:strength2}) at $F=0$
and (\ref{E:strength1}) at $\beta \ll 1$, respectively,
(see eqs. (\ref{E:link}) and (\ref{E:link1})).
Peak positions are counted with respect to the
ground $(p=0)$ 2D peak energy
$\hbar\omega_0 =\tilde{E_g} - 4Ry$.}
\end{figure}

$$
\frac{E_{00}^{(b)} - 4Ry}{4Ry}=-
\left(\frac{9 \pi^2\mu}{ 4 m_{\parallel}(0)}\right)^{\frac{1}{3}}\,\mbox{and}\,~
\frac{f_{00}}{f_{00}^{(2\mbox{\tiny D})}}=
\frac{d}{a_0}
\left(\frac{16 m_{\parallel}(0)}{\pi^2\mu}\right)^{\frac{1}{3}}
\sin^2 \frac{7\pi}{12},
$$
respectively. The oscillator strengths and the blue shifts
of the ground and some excited satellite exciton peaks versus
the electric fields $F$ are depicted in figures 2 and 3,
respectively.
With increasing electric field $F$
the $0s$ maxima oscillate, reduce
in magnitude and red-shift towards the
energy $\hbar \omega_0$.

\begin{figure}[htbp]
 \begin{center}
\includegraphics[width= .75\linewidth]{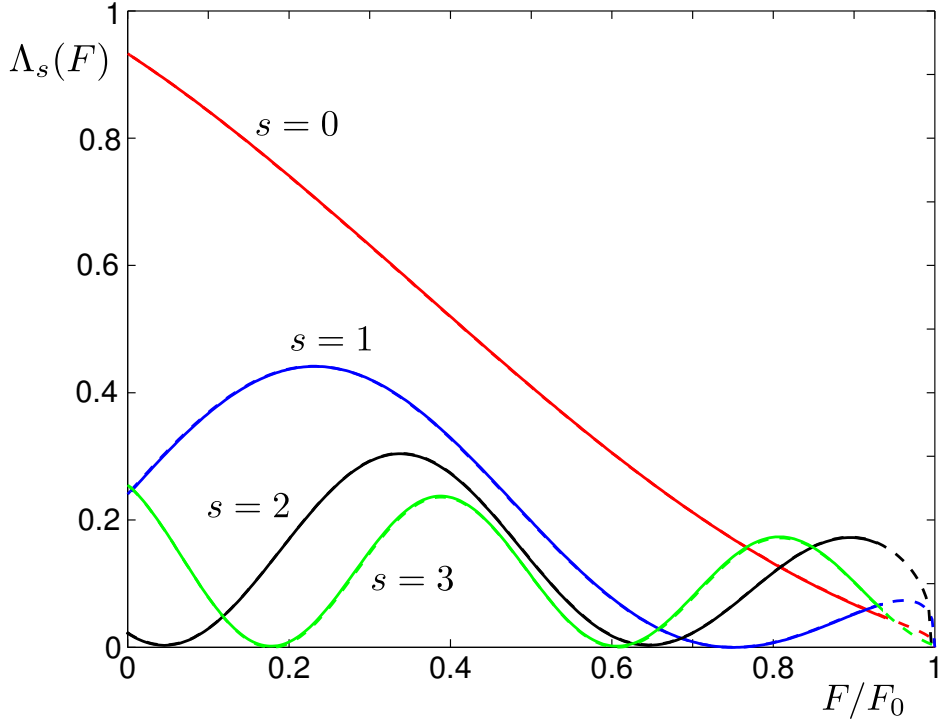}
 \end{center}
 \caption{The oscillator strengths of the ground $(f_{00})$
and excited $(f_{01},f_{02}, f_{03} )$
satellite exciton peaks scaled with the 2D oscillator
strength $f_{00}^{(2\mbox{\tiny D})}$ calculated from
eqs. (\ref{E:strength2}) and (\ref{E:strength1})
at $\beta \ll 1$,
respectively, (see eq. (\ref{E:link}) )
as a function of the relative electric fields
$\frac{F}{F_0}$.}
\end{figure}

\begin{figure}[htbp]
 \begin{center}
\includegraphics[width= .75\linewidth]{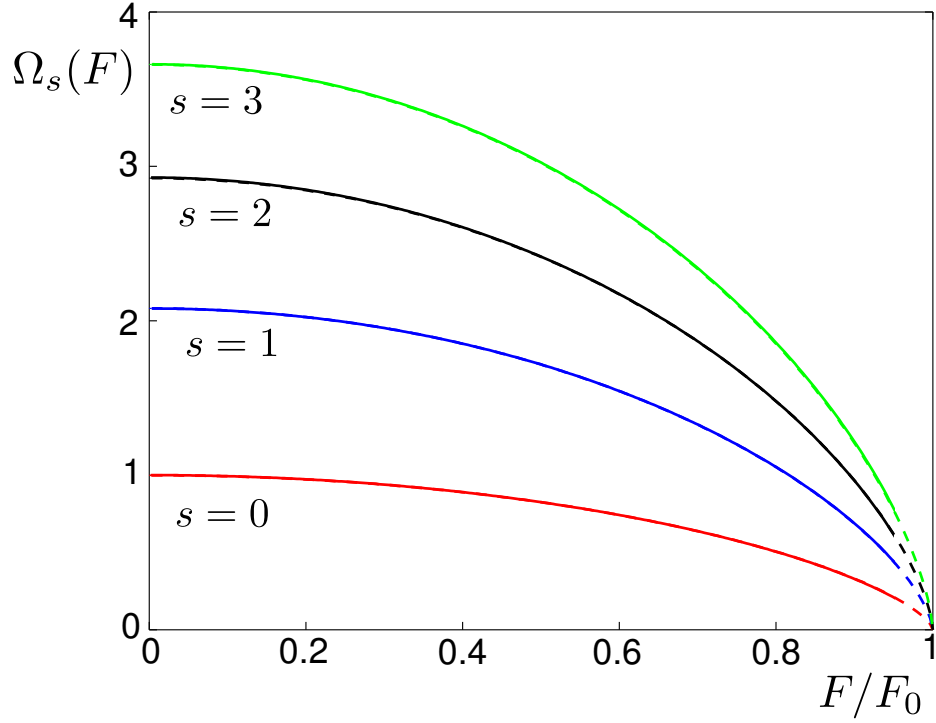}
 \end{center}
 \caption{The dependencies of the blue shifts
$\hbar\omega - (\tilde{E_g} - 4Ry)$ of the ground
$s=0$ and excited $s=1,2,3$ satellite
exciton peaks scaled with the 2D exciton
binding energy $4Ry$ (see eq. (\ref{E:link1}) )
on the relative electric fields $\frac{F}{F_0}$.
The blue shifts are calculated from eqs.
(\ref{E:spectr2}) and (\ref{E:bind})
and counted from the position of the
ground $(p=0)$ 2D peak $\hbar\omega_0 =\tilde{E_g} - 4Ry$.}
\end{figure}

 At $F\simeq F_0 $ the fine structure
of the optical spectrum
turns into a quasi-continuous subband
that in turn allows us to replace in eqs.
(\ref{E:spectr2})-(\ref{E:bind})

$\sum_s~\mbox{by}~\int\frac{ds}{d E_{0s}^{(b)}}d E_{0s}^{(b)}
~\mbox{with}~\frac{ds}{d E_{0s}^{(b)}}=
\frac{3}{16Ry}
\left[(2s+1 )\frac{m_{\parallel}(0)}{9 \pi^2 \mu}\right]^{\frac{1}{3}}
\left(1 - \frac{F}{F_0}  \right)^{-\frac{2}{3}}$.
This gives, as expected, the quasi-2D exciton absorption
in an electrically unbiased SL in the vicinity of the
bottom of the miniband branching from the
ground $(p=0)$ exciton peak
$\alpha \sim Ry (\Delta_e +\Delta_h)^{-\frac{1}{2}}
[\hbar\omega - (\tilde{E}_g - 4Ry)]^{-\frac{1}{2}}$
(see for example Ref. \cite{monzhbatdun94} ). The limiting case
$F=F_0$ is excluded by the conditions (\ref{E:final}).

The continuous regime implies the adiabatic
separation of the fast $\vec{\rho}$-motion in the heteroplanes
governed by the 2D Coulomb potential and the slow motion
parallel to the SL $z$-axis within the asymmetric triangular
quantum well. This requires the extent of the wave function
$\psi(z)$ (\ref{E:psi}) $a_{+,-}(F)$ to be much less
than the 2D exciton Bohr radius $\frac{1}{2}a_0$. Taking
into account the basic condition (\ref{E:adiab}) we obtain

\begin{equation}\label{E:final}
d\ll a_0 \left(\frac{\mu}{32 m_{\|}(0)}\right)^{\frac{1}{3}}
\left(1-\frac{F}{F_0}\right)^{-\frac{1}{3}}\ll\frac{1}{2}a_0.
\end{equation}

In addition, the effect of the longitudinal motion of the exciton
with the finite reduced
mass $m_{\parallel}(0) \sim \Delta_{eh}^{-1} d ^{-2}$ should be
a small perturbation to the 2D exciton binding energy
$4Ry$ in eq. (\ref{E:bind}) for $E_{0s}^{(b)}$ that in turn requires

\begin{equation}\label{E:final1}
\left(\frac{\mu}{m_{\|}(0)}\right)^{\frac{1}{3}}
\left(\frac{3\pi}{2}\right)^{\frac{2}{3}}
\left(1- \frac{F^2}{F_0^2}\right)^{\frac{2}{3}} \ll 1.
\end{equation}

The applicability of the obtained results needs
individual discussions. Eqs. (\ref{E:final}) and (\ref{E:final1})
impose strong restrictions on the SL parameters
$\Delta_{e,h}, d, \mu$ and the electric fields
$F$ that in turn prevents us from detailed numerical
estimates of the expected experimental values for the
concrete SLs. In particular eq. (\ref{E:final1}) implies
on the one hand a small reduced effective mass $\mu$,
narrow miniband $\Delta_{eh}$ and short SL period $d$,
while on the other hand $\Delta_{eh} \sim \mu^{-1}d^{-2}$.
However, as pointed out in Ref. \cite{suris} this
contradiction can be lifted by the tunneling factor
governed by the optimal relationship between the barrier
and well widths. This factor can be considered as
independent of the lattice parameters.
As for the electric fields
$F$ that its are on the one
hand promote the fulfillment of eq.
(\ref{E:final1}) and continual regime and on the other hand prevent
use of the adiabatic approximation both reflected in eq. (\ref{E:final}).

It follows from eq. (\ref{E:spectr2}) that the red shift
$\omega (F) - \omega (0)$ of the ground $(s=0)$ satellite peak
corresponding to the ground Rydberg series $(p=0)$
for weak electric fields $F<<F_0$ reads

$$
\omega (F) - \omega (0) = - \frac{4 Ry}{\hbar}
\left(\frac{2\pi^2\mu}{3 m_{\parallel}(0)}\right)^{\frac{1}{3}}
\left(\frac{F}{F_0}\right)^2
$$
and allows us to calculate the longitudinal effective reduced
mass $m_{\parallel}(0)$. Subsequently the width of the
total electron-hole miniband $\Delta_{eh}(0)=\Delta_{e} +\Delta_{h} $
affecting the interband electronic, optical and transport properties
of the semiconductor SLs can be derived.

Focussing on possible experiments estimates of the expected
values can be made for the parameters of the
GaAs/Al$_{x}$Ga$_{1-x}$As $(x=0.3)$ SL of period $d=3\,\mbox{nm}$
with the parameters
$\mu = 0.060 m_0,~\epsilon = 13.2,~Ry=4.7\,\mbox{meV},~a_0 = 11.5\,\mbox{nm}$
\cite{harr}. The exciton electric field determining the longitudinal
exciton potential $\bar{U}_0 (z)$ (\ref{E:1Dpot}) is $F_0 =130\,\mbox{kV/cm}$.
As pointed out above in view of the conditions
(\ref{E:final}) and (\ref{E:final1}) at $F=0$ this SL
with the realistic miniband width does not seem to be a
candidate for an experimental study. However, in the case of the miniband
width $\Delta_{eh}\simeq 30\,\mbox{meV}~\mbox{and}~F=0.88F_0$
the binding energy $E_{00}^{(b)}$ (\ref{E:bind})
becomes $E_{00}^{(b)}/4Ry \simeq 0.58$.
At the same time this electric field
considerably reduces the corresponding peak intensities to give for the
oscillator strengths (\ref{E:strength2})
$f_{00} (F)=0.062f_{00} (0)$.
Clearly, the electric fields with strengths $F\simeq F_0$
narrow down the distances between the satellite peaks
for different quantum numbers $s$ and compose the quasi-continuous
absorption band.
Note that
the terms $\sim u^2, \mid u\mid ^3$ in eq. (\ref{E:1Dpot})
reduce the effect of the potential
$\bar{U}_0 (z)$ on the 2D exciton binding energy
$4Ry$ in eq. (\ref{E:bind}) for $E_{0s}^{(b)}$,
contribute
to the condition (\ref{E:final1}) and thus facilitate the search for
suitable lattices.
We believe that the continuing
technological progress in fabricating narrow miniband short
period SLs renders our
analytical results relevant and can contribute to the
understanding of the nature of the SL exciton spectra
as well as further promote its application to optoelectronics.

\section{Conclusions}\label{S:concl}

We have derived analytically the exciton wave functions
and energies and the exciton absorption coefficient
in semiconductor superlattices in the presence of
external dc electric fields directed parallel to the SL axis.
The SL period was taken to be much smaller than the exciton
Bohr radius. Strong and weak electric fields providing
the spatially localized and extended exciton states, respectively,
were considered in the adiabatic approximation.
In weak electric fields regime we focus on the fine structure
of each of the 2D exciton energy levels grouped into the Rydberg series.
This structure is related to the bound electron-hole states
in the quasi-uniform longitudinal exciton
electric field that in turn reduces the exciton binding energy.
It was shown that the external
electric field compensates the exciton electric field,
considerably modifies the exciton states and corresponding
optical exciton fine structure and recovers the strictly 2D
exciton binding energy and the 2D exciton absorption in
an unbiased SL.

\section{Acknowledgments}\label{S:Ackn}
The authors are grateful to A. Zampetaki for valuable technical assistance.


\begin{thebibliography}{99}
\bibitem{suris}
R.~A.~Suris, Semiconductors, \textbf{49}, 807 (2015).

\bibitem{kohnlutt}
W.~Kohn, J.~M.~Luttinger, Phys. Rev. \textbf{98}, 915 (1955).

\bibitem{zhil92}
A.~G.~Zhilich, Sov. Phys. Solid. State, \textbf{34}, 1875 (1992).

\bibitem{monzhil94}
B.~S.~Monozon and A.~G.~Zhilich, Phys. Solid. State, \textbf{37}, 508 (1995).

\bibitem{monschm05}
B.~S.~Monozon and P.~Schmelcher, Phys. Rev. B \textbf{71}, 085302 (2005).

\bibitem{ivch}
S.~M.~Cao, M.~Willander, E.~L.~Ivchenko, A.~I.~Nesvizhskii and A.~A.~Toropov,
Superlattices Microstructures, \textbf{17}, 97, (1995).

\bibitem{abram}
\emph{Handbook of Mathematical Functions}, edited by M.~Abramowitz and
I.~A.~Stegun (Dover, New York, 1972).

\bibitem{ell}
R.~J.~Elliot, Phys. Rev. \textbf{108}, 1384 (1957).

\bibitem{weiler}
M.~H.~Weiler, M.~Reine, and B.~Lax, Phys. Rev. \textbf{171}, 949 (1968).

\bibitem{monzhbatdun94}
B.~S.~Monozon, A.~G.~Zhilich, C.~A.~Bates and J.~L.~Dunn,
J. Phys.: Condens. Matter \textbf{6}, 10001 (1994)

\bibitem{harr}
P.~Harrison, \emph{Quantum Wells, Wires and Dots} (Wiley, New York, 2000), p. 183.
\end{thebibliography}
\end{document}